\documentclass{article}
\usepackage{graphicx}  
\usepackage{amsmath}   
\usepackage[compress]{cite}
\usepackage{amssymb}   
\usepackage{bm} 
\usepackage{dcolumn}
\usepackage{color}
\usepackage{mathrsfs}
\usepackage{amsfonts}
\usepackage{varioref}
\usepackage{textcomp}
\RequirePackage[colorlinks,citecolor=blue,urlcolor=magenta,linkcolor=blue]{hyperref}
\allowdisplaybreaks
\labelformat{section}{Section #1} 
\labelformat{subsection}{Section #1} 
\labelformat{subsubsection}{Section #1}
\labelformat{subsubsubsection}{Section #1}
\labelformat{equation}{Eq.~(#1)} 
\labelformat{figure}{Fig.~#1} 
\labelformat{subfigure}{Fig.~\thefigure#1} 
\labelformat{table}{Tab.~#1} 
\labelformat{appendix}{Appendix #1}
\title{Embedding into flat spacetime and black hole thermodynamics} 
\author{T.R. Govindarajan
\footnote{trg@imsc.res.in}~$^{1,2}$
and Sumanta Chakraborty
\footnote{sumantac.physics@gmail.com}~$^{3}$\\
{$^{1}$\small{Chennai Mathematical Institute, Kelambakkam, Siruseri, Tamil Nadu 600113, India}}\\
{$^{2}$\small{The Institute of Mathematical Sciences, Taramani, Tamil Nadu 600113, India}}\\
{$^{3}$\small{School of Mathematical and Computational Sciences and School of Physical Sciences}}\\ 
{\small{Indian Association for the Cultivation of Science, Kolkata-700032, India}}}
\date{ \today}  
\begin{document}

\maketitle

\begin{abstract}

It is known that static and spherically symmetric black hole solutions of general relativity in different spacetimes can be embedded into higher dimensional flat spacetime. Given this result, we have explored the thermodynamic nature of black holes \'{a} la its embedding into flat spacetime. In particular, we have explicitly demonstrated that black hole temperature can indeed be determined starting from the embedding and hence mapping of the static observers in black hole spacetime to Rindler observers in flat spacetime. Furthermore, by considering the dynamics of a scalar field in the flat spacetime it is indeed possible to arrive at the area scaling law for black hole entropy. Thus using flat spacetime field theory, one can indeed provide a thermodynamic description of black holes. Implications are also discussed.      

\end{abstract}
\section{Introduction}

Black holes, so far, are the simplest and elegant solution of general relativity, harbouring several counter-intuitive features, describing gravitational interaction at various different length scales. Now that there are ample indirect evidence for the existence of black holes, from gravitational wave observations \cite{Abbott:2016blz,TheLIGOScientific:2016src} as well as the detection of black hole shadow \cite{Akiyama:2019cqa,Akiyama:2019eap}, it is important to understand the physics associated with the mysteries behind the black holes better. The first in such a list of mysteries will be the weird connection between the dynamics of a black hole to thermodynamic laws. An understanding of the illusive thermodynamic nature of gravity started ever since Bekenstein suggested that black holes have entropy \cite{Bekenstein:1972tm,Bekenstein:1973ur} and of course this has been put on a firm ground by the subsequent discovery of Hawking that black holes radiate \cite{Hawking:1974sw,Davies:1974th,Unruh:1976db}. To be precise, if there is an ambient quantum field in a black hole spacetime, then the vacuum fluctuations associated with an in-falling observer will appear as thermal fluctuations to a static observer, with a temperature inversely proportional to the mass of the black hole.   

There have been numerous attempts in the literature to explain the thermodynamical properties of black holes from more fundamental principles, possibly quantum mechanical in origin \cite{tHooft:1984kcu,Bombelli:1986rw,Carlip:1995cd}. The most common consensus regarding the celebrated area law for black hole entropy, central to black hole thermodynamics, suggests that the microscopic degrees of freedom responsible for entropy of the black hole reside on the black hole horizon and that the number of such microscopic states is proportional to the exponential of the horizon area \cite{Kaul:2000kf}. It should be emphasized that the area entropy relation is very particular to black holes in general relativity and ceases to hold in theories beyond general relativity \cite{Wald:1993nt,Iyer:1994ys,Jacobson:1993xs} (however, there are exceptions \cite{Chakraborty:2018qew}). Further, the correspondence between gravity and thermodynamics seems to be far deep rooted than originally thought \cite{Padmanabhan:2013nxa,Chakraborty:2015wma}, since Einstein's equations themselves can be expressed as thermodynamic identities on any arbitrary null surface, e.g., as a first law and Navier-Stokes equation of fluid dynamics \cite{Chakraborty:2015hna,Chakraborty:2015aja}. This demonstrates why solutions of gravitational field equations, in particular black hoes, exhibited the thermodynamic properties. Interestingly, these results transcend general relativity and holds in Lanczos-Lovelock theories of gravity as well, which once again demonstrates that gravity thermodynamics correspondence hints at microscopic ``atomic" structure of gravity \cite{Padmanabhan:2009vy,Padmanabhan:2013nxa,Chakraborty:2019vki}.  

It is more or less generally accepted that a full understanding of black hole thermodynamics, in particular the origin of black hole entropy would require knowledge of quantum theory of gravity or the microscopic structure of spacetime. Despite, the lack of a proper quantum gravity theory describing the microstructure of spacetime, in certain scenarios one has been able to calculate the area-entropy relation. These methods include --- (a) brick wall approach by t'Hooft \cite{tHooft:1996rdg}, (b) in the context of extremal/near-extremal black holes in string theory \cite{Sen:1995in,Strominger:1996sh} and (iii) of course, in the loop quantum gravity approach \cite{Rovelli:1996dv,Ashtekar:1997yu}. Even though we do get the area-entropy relation, this depends on the details of the quantum gravity model. 

Another possibility to infer the area law for black hole entropy, in the context of general relativity, without going into details of quantum gravity is to use the result that quantum entanglement between the degrees of freedom of a quantum field, that are outside with those inside the event horizon, scales as area. This can be understood along the following lines, since the degrees of freedom responsible for black hole entropy are inside the horizon and hence are inaccessible to a distant observer, they must be summed over, yielding a density matrix whose van Neumann entropy is proportional to the area of the horizon \cite{Balachandran:2013cq,Solodukhin:2011gn,Solodukhin:2008dh}. In particular, It is well known that in manifolds with boundary, the Laplacian operator has self adjoint extensions and the edge states localized at the boundary arises naturally. They may serve as models for accounting the microstates associated with a given black hole geometry \cite{Govindarajan:2011hs,Govindarajan:2015hfa,Govindarajan:2013xka}. We will explore this avenue further in this work. 

This letter is organized as follows: In \ref{trgsc_sec2} we will discuss the general formalism of embedding a static and spherically symmetric spacetime within a flat background, some examples of which will be discussed in \ref{trgsc_sec3}. Using the embedding into flat spacetime of a black hole geometry, we will discuss in \ref{trgsc_sec4} how the notion of temperature and subsequently in \ref{trgsc_sec5} how the notion of entropy comes about, leading to a thermodynamic description of black holes. Finally we conclude with a discussion on the obtained results and future prospects.  
\section{Embedding a general spherically symmetric and static spacetime into a flat Lorentzian manifold}\label{trgsc_sec2}

Minkowski spacetime is one of the most simplest and well understood spacetime, on which various physics problems can be exactly solved including the  quantum field theory. Besides there is also a very close correspondence between Minkowski spacetime and Rindler spacetime, which in turn leads to the remarkable result that with respect to a uniformly accelerated observer, the Minkowski vacuum appears thermal in nature. A very similar result holds for black hole spacetimes as well, where the initial vacuum state will appear to be thermal for a static observer at late times. Hence one may ask whether there is any close correspondence between these two scenarios and as expected the answer turned out to be yes \cite{Deser:1997ri,Jacobson:1997ux,Deser:1998xb,Padmanabhan:2002ha} (see also \cite{Banerjee:2010ma}). It follows that one can indeed embed Schwarzschild spacetime in a higher dimensional flat spacetime, where static observers in Schwarzschild spacetime maps to Rindler observers in flat higher dimensional spacetime. This provides an elegant description for thermality of black hole horizons. In this work, we would like to generalize the analysis to an arbitrary static and spherically symmetric spacetime.

Thus let us start by considering a general four dimensional spacetime which is spherically symmetric as well as static. The line element in tune with the above symmetries can be written down as,
\begin{align}\label{trgsc_eq1}
ds^{2}=-f(r)dt^{2}+\frac{dr^{2}}{g(r)}+r^{2}\left(d\theta^{2}+\sin ^{2}\theta d\phi \right)^{2}~.
\end{align}
The task is to embed the above four-dimensional line element into a higher dimensional flat spacetime. Taking a cue from the scenario in Schwarzschild spacetime, we will try to embed the four dimensional spacetime into a flat six dimensional spacetime. For this purpose we can introduce the following coordinate transformations,
\begin{align}\label{trgsc_eq2}
Z^{0}&=a\sqrt{f(r)}\sinh (t/a);\qquad Z^{1}=a\sqrt{f(r)}\cosh (t/a);\qquad Z^{2}=\int ^{r}dr \sqrt{h(r)};
\nonumber
\\
Z^{3}&=r\sin \theta \cos \phi;\qquad Z^{4}=r\sin \theta \sin \phi;\qquad Z^{5}=r\cos \theta~.
\end{align}
Here $a$ is a constant having dimension of Length, related to some characteristic length scale of the solution (this would be related to the mass of the black hole for Schwarzschild spacetime). In defining the coordinate $Z^{2}$, we have introduced a function of radial coordinate $h(r)$, which as of now is arbitrary and will be determined later. Further, the above transformation introduces the coordinates $Z^{0}$ and $Z^{1}$ outside the event horizon, while in order for them to exist inside the horizon as well, it is necessary to modify the above transformation by changing $\sqrt{f(r)}$ to $\sqrt{|f(r)|}$ and $\cosh(t/a)$ to $\sinh(t/a)$ and vice versa \cite{Banerjee:2010ma}. 

Given the above transformation, we can compute the differential of all the flat space coordinates, including $dZ^{0}$ and $dZ^{1}$ respectively, which leads to the following expression for the six dimensional flat line element,
\begin{align}\label{trgsc_eq3}
-\left(dZ^{0}\right)^{2}&+\left(dZ^{1}\right)^{2}+\left(dZ^{2}\right)^{2}+\left(dZ^{3}\right)^{2}+\left(dZ^{4}\right)^{2}+\left(dZ^{5}\right)^{2}
\nonumber
\\
&=-f(r)dt^{2}+\left(\frac{a}{2}\right)^{2}\frac{f'^{2}}{f}dr^{2}+h(r)dr^{2}+dr^{2}+r^{2}\left(d\theta^{2}+\sin ^{2}\theta d\phi ^{2}\right)
\nonumber
\\
&=-f(r)dt^{2}+\left\{\left(\frac{a}{2}\right)^{2}\frac{f'^{2}}{f}+h(r)+1\right\}dr^{2}+r^{2}\left(d\theta^{2}+\sin ^{2}\theta d\phi ^{2}\right)
\end{align}
Thus in order to match the six dimensional flat spacetime with the original four dimensional static and spherically symmetric line element, we need to fix the function $h(r)$ such that, 
\begin{align}\label{trgsc_eq4}
h(r)=\frac{1}{g(r)}-1-\left(\frac{a}{2}\right)^{2}\frac{f'^{2}}{f}
\end{align}
This provides a general rule to embed a static and spherically symmetric spacetime in a higher dimensional flat spacetime. However, in general for different $g(r)$ and $f(r)$, the function h(r) will not be continuous at the black hole horizon (located at $g(r)=0$). Thus at face value it seems that the above embedding procedure will not work. Fortunately, if one restricts to a specific class of static and spherically symmetric metric, which satisfies the condition $f(r)=g(r)$ and for a certain choice for $a$, the function $h(r)$ will be regular at the black hole horizon. 

Thus we will concentrate on the spacetime metrics having $f(r)=g(r)$ and in order to fix the constant $a$ we will explore the near horizon limit of the function $h(r)$, which following \ref{trgsc_eq4} takes the form,
\begin{align}\label{trgsc_eq5}
h(r_{\rm h}+\epsilon)=\frac{1}{f(r_{\rm h})+\epsilon f'(r_{\rm h})}\left[1-\left(\frac{a}{2}\right)^{2}\left\{f'(r_{\rm h})^{2}+2\epsilon f'(r_{\rm h})f''(r_{\rm h})+\mathcal{O}(\epsilon^{2})\right\}\right]-1
\end{align}
where $\epsilon=r-r_{\rm h}$. Since by definition $f(r_{\rm h})=0$, it follows that, the above expression will be finite, if and only if, $a\equiv (2/f'(r_{\rm h}))=\kappa _{\rm h}^{-1}$. Thus for this choice of $a$ we obtain the near horizon behaviour of the function $h(r)$ to become, $h(r_{\rm h})=-(2f''(r_{\rm h})/f'(r_{\rm h})^{2})-1$, which is certainly finite as long as $f'(r_{\rm h})$ is non-zero, or in other words the surface gravity is finite. Thus for non-extremal black holes, the above prescription will always provide embedding of a static spherically symmetric black hole into a higher dimensional flat spacetime. Note that even though we have worked with four dimensional spacetime, the same procedure will hold true for higher dimensional spacetimes as well \cite{Santos:2004ws}. In particular, if one starts from a $d$ dimensional static and spherically symmetric spacetime, it is possible to embed it in a $(d+2)$ dimensional flat spacetime. We will see an example of this in the next section.

However, a note of caution is necessary here. The above analysis of embedding a four dimensional black hole spacetime into a higher dimensional flat manifold works for a single horizon, since the continuity of the function $h(r)$, introduced above, can be ascertained only at $r_{\rm h}$. If the black hole solution has other horizons (e.g., Cauchy horizon) then the function $h(r)$ will not remain regular over there. If one needs an embedding which is regular at both the horizons, another time coordinate must be introduced \cite{Deser:1998xb,Kim:2000ct,Chen:2004qw}. This is of serious concern for asymptotically de Sitter spacetimes, since both the horizons are accessible by static observers, but for asymptotically flat spacetime, the Cauchy horizon is within the black hole event horizon and thus is not of much concern. Hence we can safely conclude that for asymptotically flat spacetimes, the above prescription works well as long as we are interested in physics outside the Cauchy horizon, which certainly makes sense as the predictability of Einstein's equations works only upto the Cauchy horizon. 
\section{Examples: Embedding into flat spacetime}\label{trgsc_sec3}

In this section, we will concentrate on three different examples in the context of general relativity along with another from pure Lovelock theories, explicitly depicting the applicability of the general method presented above. These will include the Schwarzschild spacetime, Reissner-Nordstr\"{o}m spacetime, non-rotating BTZ black hole as well as a pure Lovelock black hole. As a warm up exercise we will start by discussing the Schwarzschild spacetime. 
\subsection{Schwarzschild Spacetime}

The most well known case corresponds to the Schwarzschild spacetime, for which it is obvious that $f(r)=g(r)=1-(2GM/r)$. Here, the surface gravity associated with the event horizon at $r=2M$ becomes $\kappa=(1/4M)$ and hence the length scale associated with the solution becomes $a=4M$. Thus the function $h(r)$, in this case, becomes,
\begin{align}\label{trgsc_eq6}
h(r)&=\frac{1}{\left(1-\frac{2M}{r}\right)}-1-\left(2M\right)^{2}\frac{1}{\left(1-\frac{2M}{r}\right)}\left(\frac{2M}{r^{2}}\right)
\nonumber
\\
&=\frac{2GM}{r}\left\{1+\left(\frac{2GM}{r}\right)+\left(\frac{2GM}{r}\right)^{2} \right\}
\end{align}
The above expression yields the unknown function $h(r)$, which is regular at the black hole horizon and positive everywhere. Hence the transformation is well-behaved and one can indeed transform the four dimensional Schwarzschild spacetime to a flat six dimensional spacetime.

\subsection{Reissner-Nordstr\"{o}m Spacetime}

As the second example, we will consider a black hole with a charge, known as the Reissner-Nordstr\"{o}m spacetime. For which the metric elements are, $f(r)=g(r)=1-(2M/r)+(Q^{2}/r^{2})$. Further the surface gravity associated with the horizon radius $r_{\rm h}=M+\sqrt{M^{2}-Q^{2}}$ correspond to $\kappa_{\rm h}=(1/2)\{(2M/r_{\rm h})^{2}-(2Q^{2}/r_{\rm h}^{3})\}$ and hence one have the length scale connected to the coordinate transformation to read, $a=2\{(2M/r_{\rm h})^{2}-(2Q^{2}/r_{\rm h}^{3})\}^{-1}$. With this choice for the length scale $a$, the unknown function $h(r)$ can be determined as,
\begin{align}\label{trgsc_eq7}
h(r)&=\left(1-\frac{2M}{r}+\frac{Q^{2}}{r^{2}}\right)^{-1}-1-\left(1-\frac{2M}{r}+\frac{Q^{2}}{r^{2}}\right)^{-1}\left(\frac{2M}{r^{2}}-\frac{2Q^{2}}{r^{3}}\right)^{2}\left(\frac{2M}{r_{\rm h}^{2}}-\frac{2Q^{2}}{r_{\rm h}^{3}}\right)^{-2}
\nonumber
\\
&=\left(1-\frac{2}{x}+\frac{Q_{*}^{2}}{x^{2}}\right)^{-1}\left\{\frac{2}{x}-\frac{Q_{*}^{2}}{x^{2}}
-\left(\frac{1}{x^{2}}-\frac{Q_{*}^{2}}{x^{3}}\right)^{2}\left(\frac{1}{x_{\rm h}^{2}}-\frac{Q_{*}^{2}}{x_{\rm h}^{3}}\right)^{-2}\right\}
\end{align}
where we have defined $x=(r/M)$ and hence the location of the horizon in the rescaled coordinate becomes $x_{\rm h}=1+\sqrt{1-Q_{*}^{2}}$. Here we have defined $Q_{*}=Q/M$ and as evident from \ref{trgsc_eq7}, the term inside the curly bracket becomes identical to $\{1-(2/x_{\rm h})+(Q_{*}^{2}/x_{\rm h}^{2})\}$ and hence the function h(r) is regular there. Thus the four dimensional Reissner-Nordstr\"{o}m black hole can indeed be embedded in a flat six dimensional spacetime.

As emphasized earlier, the function $h(r)$ presented in \ref{trgsc_eq7} is regular at the black hole event horizon, but not at the Cauchy horizon. This does not pose a serious trouble as the Cauchy horizon is always clocked by an event horizon to an external observer and by strong cosmic censorship conjecture spacetime is in-extendible beyond the Cauchy horizon. Thus continuity of the embedding across the black hole event horizon will suffice for our purpose.
\subsection{Non-rotating BTZ Black Hole}

As the third example, let us consider three dimensional non-rotating BTZ black hole \cite{Banados:1992wn}. In this context we can try to embed the BTZ black hole in a five dimensional spacetime, with $Z^{0}$, $Z^{1}$ and $Z^{2}$ remaining identical to that in \ref{trgsc_eq2}, while we define $Z^{3}=r\cos \theta$ and $Z^{4}=r\sin \theta$. The metric elements associated with the non-rotating BTZ black hole has the form, $f(r)=g(r)=(r^{2}-r_{+}^{2})/\ell ^{2}$, while the mass is defined as, $M=r_{+}^{2}/\ell ^{2}$. The surface gravity associated with the horizon located at $r_{+}$ becomes, $\kappa_{\rm h}=r_{+}/\ell^{2}$ and hence we have $a=\ell^{2}/r_{+}$. Then the unknown function $h(r)$ in the context of non-rotating BTZ black hole becomes,
\begin{align}\label{trgsc_eq8}
h(r)=\frac{\ell ^{2}}{(r^{2}-r_{+}^{2})}\left[1-\frac{(r^{2}-r_{+}^{2})}{\ell ^{2}} -\frac{r^{2}}{r_{+}^{2}} \right]
\end{align}
As evident, in this case as well, $h(r)$ is regular at the event horizon $r=r_{+}$ and hence the three dimensional non-rotating BTZ black hole can be embedded in a five dimensional flat spacetime. 
\subsection{Aside: Black holes in pure Lovelock theories}

To illustrate that the prescription presented above indeed works, even in higher dimensions and for higher curvature black holes, we will consider the static and spherically symmetric black hole solution in pure Lovelock theories and it's embedding within a higher dimensional flat spacetime. We will take the pure Lovelock black hole to be a solution of the Lovelock Lagrangian of order $m$ in $d$ spacetime dimensions, for which the metric elements of the static and spherically symmetric spacetime read \cite{Gannouji:2013eka,Dadhich:2012ma},
\begin{align}\label{trgsc_eq9}
ds^{2}=-f(r)dt^{2}+\frac{dr^{2}}{f(r)}+r^{2}d\Omega_{d-2}^{2};\qquad f(r)=1-\left(\frac{2M}{r^{(d-2m-1)/m}}\right)
\end{align}
We will now try to embed this d-dimensional spacetime in a $(d+2)$ dimensional flat spacetime. This requires determination of the function $h(r)$, which for the above metric takes the following form, 
\begin{align}\label{trgsc_eq10}
h(r)&=\left\{1-\left(\frac{2M}{r^{(d-2m-1)/m}}\right)\right\}^{-1}
\nonumber
\\
&\times\left[\frac{2M}{r^{(d-2m-1)/m}}-\left(\frac{a}{2}\right)^{2} 
\left\{\frac{(d-2m-1)}{m}\frac{2M}{r^{(d-2m-1)/m}}\frac{1}{r} \right\}^{2}\right]
\end{align}
The horizon associated with the above metric is located at $r_{\rm h}=(2M)^{m/(d-2m-1)}$ and hence one can immediately determine the surface gravity associated with the horizon of the above metric element to become $\kappa_{\rm h}=(d-2m-1/2m)r_{\rm h}^{-1}$, where the relation $(2M/r_{\rm h}^{(d-2m-1)/m})=1$ has been used. Therefore, the scale $a$ appearing in the embedding will have the form $a=(2m/d-2m-1)r_{\rm h}$. Therefore, substitution of $a$ in \ref{trgsc_eq10} leads to the following expression for $h(r)$,
\begin{align}\label{trgsc_eq11}
h(r)=\left\{1-\left(\frac{2M}{r^{(d-2m-1)/m}}\right)\right\}^{-1}\left(\frac{2M}{r^{(d-2m-1)/m}}-\frac{(2M)^{2}}{r^{2(d-2m-1)/m}}\frac{r_{\rm h}^{2}}{r^{2}}\right)
\end{align}
If one evaluates the above function on the event horizon $r_{\rm h}$, it immediately follows that $h(r)$ is regular at the event horizon. In particular, for $d=3m+1$, pure Lovelock black holes can be embedded in a higher dimensional flat spacetime with an $h(r)$ identical to that of Schwarzschild. The event horizon in the higher dimensional flat spacetime can be parametrized by, $Z^{0}=Z^{1}=Z^{2}=0$ and $(Z^{4})^{2}+\cdots +(Z^{d+1})^{2}=(2M)^{2m/(d-2m-1)}$.

This finishes our discussion regarding embedding of a curved black hole spacetime to flat Minkowski spacetime, though in higher dimensions. We have discussed four different examples and have demonstrated that the general analysis presented in \ref{trgsc_sec2} indeed works and it yields regular coordinate transformation at the horizon. We have also demonstrated that this technique will work even in higher dimensions and for higher curvature theories, e.g., pure Lovelock.  
\section{Estimating Black Hole Temperature from Embedding}\label{trgsc_sec4}

In the previous section we have discussed the embedding of various static and spherically symmetric spacetimes to a higher dimensional flat spacetime. Here we will use that result to understand the notion of black hole temperature. As we will see the Rindler observer in flat spacetime will again play a very crucial role in this analysis. At this stage it is worth emphasizing that the construction of embedding of a black hole spacetime into a higher dimensional flat specatime, presented above, works for a single horizon and its neighbourhood, but not for a spacetime inheriting two horizons. But, for Reissner-Nordstr\"{o}m spacetime, the other horizon is inside the black hole horizon and hence the embedding works perfectly fine for spacetime region around the outer horizon or the black hole event horizon and \emph{not} for the full spacetime. This will be sufficient for our purpose while assessing the thermal properties of event horizon. In what follows we will demonstrate the notion of black hole temperature by three different ways, exploiting the flat nature of the higher dimensional spacetime.

\paragraph*{Using trajectory of the static observers:} The Hawking temperature associated with the static observer living outside the black hole horizon can be related to the Davies-Unruh temperature of an accelerated observer in the higher dimensional flat spacetime. The trajectory of a static observer in the exterior region of a black hole spacetime is characterized by $r=\textrm{constant}=\theta=\phi$. Thus the associated trajectory in higher dimensional flat spacetime correspond to $Z^{2}=\textrm{constant}=Z^{3}=Z^{4}=Z^{5}$, while, $(Z^{0})^{2}-(Z^{1})^{2}=-f(r)/\kappa_{\rm h}^{2}=\textrm{constant}$. Hence the trajectory of a static observer becomes a hyperbolic trajectory in the higher dimensional flat spacetime. Moreover, the region outside the horizon (with $f(r)>0$), corresponds to $|Z^{1}|>|Z^{0}|$, while the $r=\textrm{constant}$ surface inside the horizon results into a trajectory with $|Z^{0}|>|Z^{1}|$. 

It is well known that \cite{gravitation,Wald:1999vt} the trajectory of a uniformly accelerated observer correspond to $X^{2}-T^{2}=a^{-2}$, where $a$ is the magnitude of the constant acceleration and $(X,T)$ are the inertial space and time coordinate. Thus the inertial vacuum will appear as thermal to the accelerated observer with the Davies-Unruh temperature being $(a/2\pi)$. Thus the trajectory of a static observer outside the horizon (it is impossible for an observer to remain static inside or on the horizon) is actually that of an accelerated observer in the higher dimensional flat spacetime with acceleration $a=(\kappa_{\rm h}/\sqrt{f(r)})$. Thus the Davies-Unruh acceleration temperature associated with such an observer corresponds to $(\kappa_{\rm h}/2\pi)f(r)^{-1/2}$. It is clear that as $r\rightarrow \infty$, $f(r)\rightarrow 1$ and the acceleration temperature becomes identical to the Hawking temperature. On the other hand at a finite radial distance, the acceleration temperature differs from the Hawking temperature by an inverse redshift factor $\sqrt{f(r)}$. Thus the trajectory of a static observer in any black hole spacetime maps to the trajectory of a uniformly accelerated observer in the higher dimensional flat spacetime in which the black hole is embedded. Furthermore, the Unruh-Davies acceleration temperature associated with the accelerated observer is related to the black hole temperature modulo an redshift factor. This provides one possible avenue to understand the thermal properties of black holes from embedding.

\paragraph*{Using the inertial propagator:} Another hint for the thermality of the horizon can be obtained by studying the nature of the inertial propagator of a massive scalar field in the higher dimensional flat spacetime within which the black hole spacetime is embedded. As we have seen in the earlier discussion, the exterior region of the black hole spacetime can be embedded within the left quadrant of the higher dimensional Minkowski spacetime, while the region inside the horizon can be embedded to the future quadrant of the higher dimensional Minkowski spacetime. Hence the region outside the horizon corresponds to the right Rindler wedge and the region inside the horizon is the future wedge. Again we will keep in mind the fact that a static observer in the region outside the horizon maps to a uniformly accelerated observer in the right wedge of the higher dimensional flat spacetime with acceleration $(\kappa_{\rm h}/\sqrt{f(r)})$. Thus we want to find out the propagator associated with the propagation between two points on the right wedge or between two points, one on the right and another on the future wedge. 

To see how the propagators can give us an estimate of the thermality associated with the black hole horizon, we write down the propagator and its Fourier transform in flat spacetime below \cite{Padmanabhan:2019yyg,Rajeev:2019bzv},
\begin{align}\label{trgsc_eq12}
G(\tau;\mathbf{x}_{1},\mathbf{x}_{2})&=i\left(\frac{1}{4\pi i}\right)^{D/2}\int^{\infty}_{0}\frac{ds}{s^{D/2}}\exp[-ism^{2}-\frac{i}{4s}\sigma^{2}(\tau;\mathbf{x}_{1},\mathbf{x}_{2})]
\nonumber
\\
A(\Omega;\mathbf{x}_{1},\mathbf{x}_{2})&=\int _{-\infty}^{\infty}d\tau G(\tau;\mathbf{x}_{1},\mathbf{x}_{2})e^{i\Omega \tau}
\end{align}
where $m^{2}$ is the mass of the scalar field and $\sigma^{2}$ is the geodesic distance. In the above Fourier transform the frequency $\Omega$ can be taken to be positive, while $A(-\Omega)$ is obtained by relating $G(\tau)$ with $G(-\tau)$. As evident, in flat, inertial coordinates the geodesic distance between any two events is simply $-(t_{2}-t_{1})^{2}+|\mathbf{x}_{2}-\mathbf{x}_{1}|^{2}$, which is symmetric under $\Delta t=t_{2}-t_{1}\rightarrow -\Delta t$ and hence it immediately follows that $A(\Omega)=A(-\Omega)$. 

We can now use the transformation of the inertial coordinates to the coordinates specific to the right wedge as well as the future wedge. If we consider the propagator between two points in the right wedge alone, the transformation to the Rindler like coordinates with $(\kappa_{h}/\sqrt{f(r)})$ as the acceleration will lead to $A(\Omega)=A(-\Omega)$. On the other hand, if we consider the propagation between right and future wedge, due to different Rindler coordinates in these two different patches, the propagator is not symmetric under time reversal and as a consequence, $A(\Omega)\neq A(-\Omega)$, with,
\begin{align}\label{trgsc_eq13} 
\frac{|A(\Omega)|^{2}}{|A(-\Omega)|^{2}}=\exp[-2\pi \Omega (\sqrt{f(r)}/\kappa_{h})]
\end{align}
If we interpret the magnitude of the Fourier transform $|A(\Omega)|^{2}$ as the probability for having a particle propagating at energy level $\Omega$, then the above result shows that the ratio of such probabilities follow a thermal distribution with a temperature $(\kappa_{\rm h}/2\pi)f(r)^{-1/2}$. This provides yet another verification of the thermal nature of black holes with an identical redshift factor, which becomes unity at large distance leading to the Hawking temperature, $(\kappa_{\rm h}/2\pi)$.  
\\
\paragraph*{Through the group of affine transformation:} There exist another method in which an estimate of the black hole temperature through its embedding onto flat spacetime can be derived. This is via the group of affine transformation, i.e., one dimensional group of translations and dilation along a real line \cite{Arzano:2018oby}. This group can also be understood as being due to two individual components, one due to translations along the real line, while the other one is dilation, but on one half of the real line, either positive/negative. Interestingly, these two operators do not commute and hence the eigenstates of these two operators are different. We will concentrate on the right half of the real line, i.e., along which the real numbers are positive. In that context it turns out that both these operators have the following position representation, $P=-i(d/dx)$ and $R=ix(d/dx)$ respectively. Since the generators do not commute it is possible to express eigenstate of one oscillator in terms of the eigenstates of another oscillator. Thus one can introduce a complete set of mode functions with respect to each of these operators and hence define a vacuum state, which with respect to the other operator will involve both right and left moving modes and hence will contain particles. The result of such an analysis is again the production of thermal particles with temperature set by the acceleration of the particles confined to the positive values of $x$. In this context as well, a static observer will be confined to the right hand wedge of the higher dimensional Minkowski spacetime and thus it divides the real line $Z^{0}=\textrm{constant}$ into two halves, $Z^{1}>Z^{0}$ and $Z^{1}<Z^{0}$. The static observers in the black hole spacetime exterior to the horizon are all located on the region $Z^{1}>Z^{0}$ and hence the above result implies that the vacuum of the full spacetime is not perceived as a vacuum on the region $Z^{1}>Z^{0}$, leading to a thermal nature for the static observers. Thus the above construction through the affine group again results into a thermal spectrum with temperature $(\kappa_{\rm h}/\sqrt{f(r)})$. This provides yet another perspective on the thermal nature of black holes.
 
\section{Klein-Gordon Equation in Higher Dimensions, Boundary Conditions and Entropy}\label{trgsc_sec5}

In the previous section we have elaborated how one may infer the existence of black hole temperature by considering physics of flat spacetime in which the black hole is embedded into. Essentially, this boils down to the fact that the black hole horizon maps to the Rindler horizon in higher dimensional flat spacetime in which the black hole is embedded into, thus the thermality of the Rindler horizon leads to the thermal nature of black holes. It would be interesting if a notion of black hole entropy can also be obtained following a similar route. Since we do not have a handle on the quantum degrees of freedom associated with gravitational interaction, the discussion presented here will be semi-classical, i.e., we will look for edge states of a quantum scalar field living on the spacetime and whether the allowed quantum states can account for the area law of black hole entropy. Here we will first present the analysis in higher dimensional flat spacetime and subsequently will discuss the corresponding situation in the black hole spacetime itself. 

For generality, we will present our result for a $d$-dimensional static and spherically symmetric black hole, inhibiting an event horizon, so that use of a single time coordinate allows one to embed it into a $(d+2)$-dimensional flat spacetime. The field equation for a scalar field in such a $(d+2)$-dimensional flat spacetime is given by the Klein-Gordan equation, which has the simple structure $\square \Phi -m^{2}\Phi=0$. This can be separated into angular part and radial part, while the time dependence is given by $e^{-i\omega t}$. Further, the black hole horizon at a given instant of time correspond to fixed values of $Z^{0}$, $Z^{1}$ and $Z^{2}$, while we have $(Z^{3})^{2}+(Z^{4})^{2}+(Z^{5})^{2}=(2M)^{2}$. Thus if we are interested in the field configuration outside the black hole horizon, we have to solve the spatial three-dimensional part of the Klein-Gordan equation in the region $\mathbb{R}^{3}-\mathbb{B}$, where $\mathbb{B}$ corresponds to the region inside the black hole horizon in the embedded flat spacetime. Thus the number of edge states associated with the above region $\mathbb{R}^{3}-\mathbb{B}$, pertaining to the hermiticity of the operator and Robin boundary condition is proportional to the two-dimensional area of the horizon \cite{Govindarajan:2011hs}. Thus one naturally ends up with the entropy area relation for black holes. It is also possible to arrive at this result by taking some other root, e.g., use of the replica trick and thus tracing over the portion of flat spacetime un-accessible to static observers outside the black hole horizon. This will also lead to an area law for entanglement entropy, known from earlier literatures \cite{Solodukhin:2011gn}. It is expected that this result in flat spacetime will transcend to the black hole spacetime, but in what follows, for completeness we would like to make another argument in favour of the entropy-area relation, but from the black hole spacetime itself.

In such a d-dimensional spacetime, the metric in the $(t,r)$ sector is given by $-f(r)dt^{2}+(dr^{2}/f(r))$. The coefficient of $dt^{2}$ is exactly the inverse of the coefficient of $dr^{2}$, since we need regular behaviour of the embedding at the event horizon. The angular sector, on the other hand, has a topology of $S^{d-2}$ and will be covered by a set of $(d-2)$ angular coordinates, $\theta_{1}, \theta _{2}, \cdots, \theta_{d-3}, \phi$, where $\theta_{i}\in (0,\pi)$ and $\phi \in (0,2\pi)$. The determinant of the full d-dimensional metric takes the form, $\sqrt{-g}=r^{(d-2)}\Omega_{d-2}(\theta_{1}, \theta_{2}, \cdots, \theta_{d-3})$. In this spacetime the Klein-Gordan equation satisfied by a massive scalar field can be expressed as,  
\begin{align}\label{trgsc_eq14}
m^{2}\Phi&=\square \Phi \equiv \frac{1}{\sqrt{-\textrm{det.}\left(g_{\mu \nu}\right)}}
\partial _{\mu}\left\{\sqrt{-\textrm{det.}\left(g_{\mu \nu}\right)}~g^{\mu \nu}\partial _{\nu}\Phi\right\}
\nonumber
\\
&=\frac{1}{r^{d-2}}\partial _{t}\left\{\left(-\frac{1}{f}\right)r^{d-2}\partial _{t}\Phi \right\}
+\frac{1}{r^{d-2}}\partial _{r}\left\{r^{d-2} g\partial _{r}\Phi \right\}
-\frac{\ell(\ell+1)}{r^{2}}\Phi
\end{align}
where, $\Phi=\Phi(t,r)$ and the angular part has been expanded into an appropriate spherical harmonic basis, leading to an overall additional factor of $\{-\ell(\ell+1)/r^{2}\}$. Since the equation has no explicit dependance on time and for future convenience, one can write down $\Phi(t,r)=e^{i\omega t}R(r)r^{-(d-2)/2}$, with $\omega>0$. At this stage it is advantageous to introduce tortoise coordinates, which is defined as $dr^{*}=dr/f(r)$. This leads to the following differential equation satisfied by the radial part of the scalar field $R(r^{*})$,
\begin{align}\label{trgsc_eq15}
-\frac{d^{2}R}{dr*^{2}}+V(r)R=\omega ^{2}R
\end{align}
which resembles the time independent Schr\"{o}dinger equation, where the potential $V(r)$, appearing in the above equation takes the following form,
\begin{align}\label{trgsc_eq16}
V(r)=m^{2}f(r)+\frac{\ell(\ell+1)}{r^{2}}f(r)+\left(\frac{d-2}{2}\right)\frac{f}{r^{(d-2)/2}}\frac{d}{dr}\left(fr^{(d-4)/2}\right)
\end{align}
The entropy associated with the above solution is intimately connected with the boundary conditions that the scalar field must satisfy, which will possibly lead to discretized quantum states. It turns out that the differential operator presented in \ref{trgsc_eq15} is self-adjoint for the Robin boundary condition, $\alpha R(r^{*})=dR(r^{*})/dr$. Here $\alpha$ is a constant with dimension of $(\textrm{Length})^{-1}$ and using appropriate limit, e.g., $\alpha \rightarrow 0$ ($\alpha ^{-1}\rightarrow \infty$) one arrives at Neumann (Dirichlet) boundary condition. One can use the definition of the tortoise coordinate in order to obtain, $\partial _{r^{*}}R=\partial _{r}R \times (dr/dr^{*})=f\partial _{r}R$. For the existence of an event horizon, it is necessary that, $f(r)$ should inherit zero at a single value of the radial coordinate $r_{\rm h}$. Then, we have $f(r)\sim (r-r_{\rm h})h(r_{\rm h})$. Therefore the Robin boundary condition translates into,
\begin{align}\label{trgsc_eq17}
A(r_{\rm h},\alpha)R(r_{\rm h})+(r-r_{\rm h})\frac{dR}{dr}\Big\vert_{r_{\rm h}}=0
\end{align}
where again, $A(r_{\rm h},\alpha)$ is a constant depending on black hole parameters through the horizon radius and the parameter appearing in Robin boundary condition. Thus there will definitely be some discrete energy states, whose number will have a maximum bound corresponding to some value $\propto \alpha r_{\rm h}$. Thus the entropy associated with the states of the scalar field living inside the horizon, scales as $n_{\rm max}^{(d-2)}$, as we sum over all possible bound states, which in turn scales as $\alpha^{2}r_{\rm h}^{(d-2)}\propto \textrm{Area}$. Note that the parameter $\alpha^{2}$ appearing in the expression for entropy can be related to the inverse of the Planck length, since this is the only meaningful length unit one can construct, we get the entropy to be $\propto (A/G_{\rm N})$. Note that the proportionality factor can not be determined by this route. Furthermore, the imposition of Robin boundary condition can be thought of along the following lines, one may interpret that the horizon has a penetration depth of the order of Planck length, which may signify a membrane fluid living near the black hole horizon. 

Thus if one now imposes the Robin boundary condition on the horizon, the density of states inside the black hole horizon will definitely scale as area density, with a proportionality factor related to the only non-trivial length scale associated with the problem, namely the Planck length. Thus from both the flat spacetime as well as curved spacetime perspective, one can infer the entropy area relation for black holes using a quantum scalar field in the respective background.

\section{Concluding Remarks}  

There are several key points which we have addressed in this work, including a thermodynamic description of black holes using flat spacetime physics! The fact that the thermodynamic behaviour is linked to a black hole spacetime comes from the embedding. This in turn relates various observers and surfaces responsible for thermodynamic description in the black hole spacetime to a similar class of observers and surfaces in flat spacetime. In particular, static observers in black hole spacetime translates to Rindler observers in the higher dimensional flat embedding. Thus the black hole temperature can be mapped to Rindler temperature associated with the embedded flat spacetime. Similarly the black hole horizon translates to a special compact surface in the flat spacetime, such that defining an appropriate hermitian operator in the region exterior to the horizon leads to edge states, whose number scales with area. Hence one can ascribe both temperature and entropy and hence argue regarding the thermodynamic nature of black holes using the fact that it can be embedded in a higher dimensional flat manifold. In this respect our work differs considerably from the earlier literatures, where main emphasize were on the embedding with one or more than one horizon and the equivalence of black hole temperature with Rindler temperature in flat spacetime. 

To summarize, elaborating on the results presented above, in this work we have shown that most of the asymptotically flat static and spherically symmetric black hole solutions can be embedded in a higher dimensional flat spacetime. This can be achieved if one considers the existence of a single horizon, around which the embedding can be well defined. In case there are two horizons, e.g., in the case of Reissner-Nordstr\"{o}m black hole, the procedure will only work for the outer event horizon, while for Cauchy horizon, the embedding will become singular. However for our purpose, in order to establish the temperature and entropy associated with the horizon, the embedding around the event horizon proved to be sufficient. Moreover, using three different procedures, namely mapping of observers, using Green's function and finally the affine group, we have defined the notion of temperature associated with the Rindler horizon in flat spacetime, which relates to the black hole temperature measured by static observers in the black hole spacetime. Similarly, by counting possible bound states associated with a scalar field in the black hole as well as in the flat spacetime, outside the horizon, we have demonstrated that the entropy indeed scales with area. Thus one can set up an analog of black hole thermodynamics in flat spacetime within which the black hole is embedded.              

In this work, we have concentrated mainly on the correspondence between flat space embedding of a spherically symmetric black hole and the associated thermodynamic properties. A similar story should hold true for rotating black holes as well, which we have not explored in this work. Further, whether some comment regarding the inner Cauchy horizon can also be made has not been probed. These we leave for the future.
\section*{Acknowledgements}

Research of S.C. is funded by the INSPIRE Faculty Fellowship (Reg. No. DST/INSPIRE/04/2018/000893) from the Department of Science and Technology, Government of India. 
\appendix

\section{Klein-Gordon Equation in Static and Spherically Symmetric Spacetime}

In this appendix we will derive the results associated with Klein-Gordan equation in a static and spherically symmetric spacetime, which will be used in the main text. In a d-dimensional static and spherically symmetric spacetime, in which we will have the $(t,r)$ sector as in the previous scenario given by $-f(r)dt^{2}+(dr^{2}/g(r))$. However, the angular sector will be a function of $\theta_{1}, \theta _{2}, \cdots, \theta_{d-3}, \phi$. Then we have, 
\begin{align}
\sqrt{-\textrm{det.}\left(g_{\mu \nu}\right)}=r^{d-2}\sqrt{\frac{f}{g}}\Omega(\theta_{1}, \theta_{2}, \cdots, \theta_{d-3})
\end{align}
Thus the Klein-Gordon equation for the radial sector takes the form,
\begin{align}
m^{2}\Phi&=\square \Phi \equiv \frac{1}{\sqrt{-\textrm{det.}\left(g_{\mu \nu}\right)}}
\partial _{\mu}\left\{\sqrt{-\textrm{det.}\left(g_{\mu \nu}\right)}~g^{\mu \nu}\partial _{\nu}\Phi\right\}
\nonumber
\\
&=\frac{1}{r^{d-2}}\sqrt{\frac{g}{f}}\partial _{t}\left\{\left(-\frac{1}{f}\right)r^{d-2}\sqrt{\frac{f}{g}}~\partial _{t}\Phi \right\}
+\frac{1}{r^{d-2}}\sqrt{\frac{g}{f}}\partial _{r}\left\{r^{d-2} g\sqrt{\frac{f}{g}}~\partial _{r}\Phi \right\}
\nonumber
\\
&\qquad \qquad \qquad \qquad-\frac{\ell(\ell+1)}{r^{2}}\Phi
\end{align}
where, $\Phi=\Phi(t,r)$ and the angular part has been separated leading to an overall additional factor of $\{-\ell(\ell+1)/r^{2}\}$. One can now write down $\Phi(t,r)=e^{i\omega t}R(r)r^{-(d-2)/2}$, leading to the following differential equation for $R(r)$,
\begin{align}
m^{2}R&=\frac{\omega^{2}}{f}R(r)-\frac{\ell(\ell+1)}{r^{2}}R(r)
+\frac{1}{r^{(d-2)/2}}\sqrt{\frac{g}{f}}\dfrac{d}{dr}\left\{r^{d-2} \sqrt{fg}~\dfrac{d}{dr}\left(r^{-(d-2)/2}R\right) \right\}
\nonumber
\\
&=\frac{\omega^{2}}{f}R(r)-\frac{\ell(\ell+1)}{r^{2}}R(r)
\nonumber
\\
&+\frac{1}{r^{(d-2)/2}}\sqrt{\frac{g}{f}}\dfrac{d}{dr}\left\{r^{(d-2)/2} \sqrt{fg}~\dfrac{dR(r)}{dr}-\left(\frac{d-2}{2}\right)\sqrt{fg}~R(r)r^{(d-4)/2} \right\}
\end{align}
Let us introduce tortoise coordinates, which is defined as $dr^{*}=dr/\sqrt{fg}$. This results into the following expression,
\begin{align}
m^{2}R&=\frac{\omega^{2}}{f}R(r)-\frac{\ell(\ell+1)}{r^{2}}R(r)
\nonumber
\\
&+\frac{1}{r^{(d-2)/2}}\sqrt{\frac{g}{f}}\frac{1}{\sqrt{fg}}\dfrac{d}{dr^{*}}\left\{r^{(d-2)/2} \frac{dR(r)}{dr^{*}}-\left(\frac{d-2}{2}\right)\sqrt{fg}~R(r)r^{(d-4)/2} \right\}
\nonumber
\\
&=\frac{\omega^{2}}{f}R(r)-\frac{\ell(\ell+1)}{r^{2}}R(r)
+\frac{1}{f}\dfrac{d^{2}R}{dr*^{2}}-\left(\frac{d-2}{2}\right)\sqrt{\frac{g}{f}}\frac{R(r)}{r^{(d-2)/2}}\frac{d}{dr}\left(r^{(d-4)/2}\sqrt{fg}\right)
\end{align}
Multiplying the above equation throughout by $f(r)$, we can rewrite the above equation as,
\begin{align}
-\frac{d^{2}R}{dr*^{2}}+V(r)R=\omega ^{2}R
\end{align}
where the potential takes the following form,
\begin{align}
V(r)=m^{2}f(r)+\frac{\ell(\ell+1)}{r^{2}}f(r)+\left(\frac{d-2}{2}\right)\frac{\sqrt{fg}}{r^{(d-2)/2}}\frac{d}{dr}\left(\sqrt{fg}r^{(d-4)/2}\right)
\end{align}
This is the expression used in the main text for the effective potential associated with the radial part of a scalar field living on the d-dimensional static, spherically symmetric spacetime. To check the correctness and validity of this result we concentrate on two examples, one for non-rotating BTZ black hole in three-dimensions and another is Schwarzschild black hole in four dimensions. 

\subsection{Example: Non-rotating BTZ Black Hole}

In this section we will apply the method presented above to determine the effective potential associated with a scalar field in a non-rotating BTZ black hole spacetime. For this spacetime, we have $f(r)=g(r)=(r^{2}-r_{+}^{2})/\ell^{2}$. Thus the potential takes the following form,
\begin{align}
V(r)&=\frac{n^{2}}{r^{2}}\left(\frac{r^{2}-r_{+}^{2}}{\ell ^{2}}\right)+m^{2}\left(\frac{r^{2}-r_{+}^{2}}{\ell ^{2}}\right)+\frac{1}{2\sqrt{r}}\left(\frac{r^{2}-r_{+}^{2}}{\ell ^{2}}\right)\frac{d}{dr}\left(\frac{r^{2}-r_{+}^{2}}{\sqrt{r}\ell ^{2}}\right)
\nonumber
\\
&=\frac{n^{2}}{\ell^{2}}-M\frac{n^{2}}{r^{2}}+m^{2}\frac{r^{2}}{\ell^{2}}-m^{2}M+\left(\frac{r^{2}-r_{+}^{2}}{\ell ^{4}}\right)
-\frac{1}{4r^{2}}\left(\frac{r^{2}-r_{+}^{2}}{\ell ^{2}}\right)^{2}
\nonumber
\\
&=\frac{n^{2}}{\ell^{2}}-M\frac{n^{2}}{r^{2}}+m^{2}\frac{r^{2}}{\ell^{2}}-m^{2}M+\frac{3}{4}\frac{r^{2}}{\ell^{4}}
-\frac{1}{2}\frac{r_{+}^{2}}{\ell^{4}}-\frac{1}{4}\frac{r_{+}^{4}}{\ell ^{4}r^{2}}
\nonumber
\\
&=\frac{n^{2}}{\ell^{2}}-M\frac{n^{2}}{r^{2}}+m^{2}\frac{r^{2}}{\ell^{2}}-m^{2}M+\frac{3}{4}\frac{r^{2}}{\ell^{4}}
-\frac{1}{2}\frac{M}{\ell^{2}}-\frac{1}{4}\frac{M^{2}}{r^{2}}
\end{align}
This matches exactly with earlier results in the literature \cite{Govindarajan:2011hs}.
\subsection{Example: Schwarzschild Spacetime}

Let us apply the above analysis to Schwarzschild spacetime, where $f(r)=g(r)=1-(2M/r)$. Thus the potential described above takes the following form,
\begin{align}
V(r)&=m^{2}\left(1-\frac{2M}{r}\right)+\frac{\ell(\ell+1)}{r^{2}}\left(1-\frac{2M}{r}\right)+\frac{1}{r}\left(1-\frac{2M}{r}\right)\frac{d}{dr}\left(1-\frac{2M}{r}\right)
\nonumber
\\
&=m^{2}\left(1-\frac{2M}{r}\right)+\frac{\ell(\ell+1)}{r^{2}}\left(1-\frac{2M}{r}\right)+\frac{2M}{r^{3}}-\frac{4M^{2}}{r^{4}}
\end{align}
In this case as well one may apply appropriate boundary conditions associated with Schr\"{o}dinger-like equation in the tortoise coordinates and hence one can get black hole entropy. Otherwise, one can start from six-dimensional flat spacetime in which the Schwarzschild black hole has been embedded. In which case the horizon is located at $\left(Z^{3}\right)^{2}+\left(Z^{4}\right)^{2}+\left(Z^{5}\right)^{2}=(2M)^{2}$, with $Z^{0}=0=Z^{1}=Z^{2}$. This is like a sphere whose interior is the region $r<2M$ and exterior is $r>2M$. Thus the interior region corresponds to $R^{3}-S^{2}$. Thus number of states on the boundary correspond to $r_{\rm h}^{2}$ \cite{Govindarajan:2011hs}, thus area-entropy relation does hold.   

\bibliography{References}

\providecommand{\href}[2]{#2}\begingroup\raggedright\begin{thebibliography}{10}

\bibitem{Abbott:2016blz}
{\bfseries Virgo, LIGO Scientific} Collaboration, B.~P. Abbott {\em et~al.},
  ``{Observation of Gravitational Waves from a Binary Black Hole Merger},''
  \href{http://dx.doi.org/10.1103/PhysRevLett.116.061102}{{\em Phys. Rev.
  Lett.} {\bfseries 116} no.~6, (2016) 061102},
\href{http://arxiv.org/abs/1602.03837}{{\ttfamily arXiv:1602.03837 [gr-qc]}}.

\bibitem{TheLIGOScientific:2016src}
{\bfseries Virgo, LIGO Scientific} Collaboration, B.~P. Abbott {\em et~al.},
  ``{Tests of general relativity with GW150914},''
  \href{http://dx.doi.org/10.1103/PhysRevLett.116.221101}{{\em Phys. Rev.
  Lett.} {\bfseries 116} no.~22, (2016) 221101},
\href{http://arxiv.org/abs/1602.03841}{{\ttfamily arXiv:1602.03841 [gr-qc]}}.

\bibitem{Akiyama:2019cqa}
{\bfseries Event Horizon Telescope} Collaboration, K.~Akiyama {\em et~al.},
  ``{First M87 Event Horizon Telescope Results. I. The Shadow of the
  Supermassive Black Hole},''
  \href{http://dx.doi.org/10.3847/2041-8213/ab0ec7}{{\em Astrophys. J.}
  {\bfseries 875} no.~1, (2019) L1},
\href{http://arxiv.org/abs/1906.11238}{{\ttfamily arXiv:1906.11238
  [astro-ph.GA]}}.

\bibitem{Akiyama:2019eap}
{\bfseries Event Horizon Telescope} Collaboration, K.~Akiyama {\em et~al.},
  ``{First M87 Event Horizon Telescope Results. VI. The Shadow and Mass of the
  Central Black Hole},'' \href{http://dx.doi.org/10.3847/2041-8213/ab1141}{{\em
  Astrophys. J.} {\bfseries 875} no.~1, (2019) L6},
\href{http://arxiv.org/abs/1906.11243}{{\ttfamily arXiv:1906.11243
  [astro-ph.GA]}}.

\bibitem{Bekenstein:1972tm}
J.~Bekenstein, ``{Black holes and the second law},''
{\em Lett. Nuovo Cimento Soc. Ital. Fis.} {\bfseries 4} (1972) 737--740.

\bibitem{Bekenstein:1973ur}
J.~D. Bekenstein, ``{Black holes and entropy},''
\href{http://dx.doi.org/10.1103/PhysRevD.7.2333}{{\em Phys.Rev.} {\bfseries D7}
  (1973) 2333--2346}.

\bibitem{Hawking:1974sw}
S.~Hawking, ``{Particle Creation by Black Holes},''
\href{http://dx.doi.org/10.1007/BF02345020}{{\em Commun.Math.Phys.} {\bfseries
  43} (1975) 199--220}.

\bibitem{Davies:1974th}
P.~Davies, ``{Scalar particle production in Schwarzschild and Rindler
  metrics},''
\href{http://dx.doi.org/10.1088/0305-4470/8/4/022}{{\em J.Phys.} {\bfseries A8}
  (1975) 609--616}.

\bibitem{Unruh:1976db}
W.~Unruh, ``{Notes on black hole evaporation},''
\href{http://dx.doi.org/10.1103/PhysRevD.14.870}{{\em Phys.Rev.} {\bfseries
  D14} (1976) 870}.

\bibitem{tHooft:1984kcu}
G.~'t~Hooft, ``{On the Quantum Structure of a Black Hole},''
\href{http://dx.doi.org/10.1016/0550-3213(85)90418-3}{{\em Nucl. Phys.}
  {\bfseries B256} (1985) 727--745}.

\bibitem{Bombelli:1986rw}
L.~Bombelli, R.~K. Koul, J.~Lee, and R.~D. Sorkin, ``{A Quantum Source of
  Entropy for Black Holes},''
\href{http://dx.doi.org/10.1103/PhysRevD.34.373}{{\em Phys. Rev.} {\bfseries
  D34} (1986) 373--383}.

\bibitem{Carlip:1995cd}
S.~Carlip, ``{Statistical mechanics and black hole entropy},''
\href{http://arxiv.org/abs/gr-qc/9509024}{{\ttfamily arXiv:gr-qc/9509024
  [gr-qc]}}.

\bibitem{Kaul:2000kf}
R.~K. Kaul and P.~Majumdar, ``{Logarithmic correction to the Bekenstein-Hawking
  entropy},'' \href{http://dx.doi.org/10.1103/PhysRevLett.84.5255}{{\em Phys.
  Rev. Lett.} {\bfseries 84} (2000) 5255--5257},
\href{http://arxiv.org/abs/gr-qc/0002040}{{\ttfamily arXiv:gr-qc/0002040
  [gr-qc]}}.

\bibitem{Wald:1993nt}
R.~M. Wald, ``{Black hole entropy is the Noether charge},''
  \href{http://dx.doi.org/10.1103/PhysRevD.48.R3427}{{\em Phys. Rev.}
  {\bfseries D48} no.~8, (1993) R3427--R3431},
\href{http://arxiv.org/abs/gr-qc/9307038}{{\ttfamily arXiv:gr-qc/9307038
  [gr-qc]}}.

\bibitem{Iyer:1994ys}
V.~Iyer and R.~M. Wald, ``{Some properties of Noether charge and a proposal for
  dynamical black hole entropy},''
  \href{http://dx.doi.org/10.1103/PhysRevD.50.846}{{\em Phys. Rev.} {\bfseries
  D50} (1994) 846--864},
\href{http://arxiv.org/abs/gr-qc/9403028}{{\ttfamily arXiv:gr-qc/9403028
  [gr-qc]}}.

\bibitem{Jacobson:1993xs}
T.~Jacobson and R.~C. Myers, ``{Black hole entropy and higher curvature
  interactions},'' \href{http://dx.doi.org/10.1103/PhysRevLett.70.3684}{{\em
  Phys. Rev. Lett.} {\bfseries 70} (1993) 3684--3687},
\href{http://arxiv.org/abs/hep-th/9305016}{{\ttfamily arXiv:hep-th/9305016
  [hep-th]}}.

\bibitem{Chakraborty:2018qew}
S.~Chakraborty and R.~Dey, ``{Noether Current, Black Hole Entropy and Spacetime
  Torsion},'' \href{http://dx.doi.org/10.1016/j.physletb.2018.10.027}{{\em
  Phys. Lett.} {\bfseries B786} (2018) 432--441},
\href{http://arxiv.org/abs/1806.05840}{{\ttfamily arXiv:1806.05840 [gr-qc]}}.

\bibitem{Padmanabhan:2013nxa}
T.~Padmanabhan, ``{General Relativity from a Thermodynamic Perspective},'' {\em
  Gen.Rel.Grav.} {\bfseries 46} (2014) 1673,
\href{http://arxiv.org/abs/1312.3253}{{\ttfamily arXiv:1312.3253 [gr-qc]}}.

\bibitem{Chakraborty:2015wma}
S.~Chakraborty, ``{Lanczos-Lovelock gravity from a thermodynamic
  perspective},'' \href{http://dx.doi.org/10.1007/JHEP08(2015)029}{{\em JHEP}
  {\bfseries 08} (2015) 029},
\href{http://arxiv.org/abs/1505.07272}{{\ttfamily arXiv:1505.07272 [gr-qc]}}.

\bibitem{Chakraborty:2015hna}
S.~Chakraborty and T.~Padmanabhan, ``{Thermodynamical interpretation of the
  geometrical variables associated with null surfaces},''
  \href{http://dx.doi.org/10.1103/PhysRevD.92.104011}{{\em Phys. Rev.}
  {\bfseries D92} no.~10, (2015) 104011},
\href{http://arxiv.org/abs/1508.04060}{{\ttfamily arXiv:1508.04060 [gr-qc]}}.

\bibitem{Chakraborty:2015aja}
S.~Chakraborty, K.~Parattu, and T.~Padmanabhan, ``{Gravitational field
  equations near an arbitrary null surface expressed as a thermodynamic
  identity},'' \href{http://dx.doi.org/10.1007/JHEP10(2015)097}{{\em JHEP}
  {\bfseries 10} (2015) 097},
\href{http://arxiv.org/abs/1505.05297}{{\ttfamily arXiv:1505.05297 [gr-qc]}}.

\bibitem{Padmanabhan:2009vy}
T.~Padmanabhan, ``{Thermodynamical Aspects of Gravity: New insights},''
  \href{http://dx.doi.org/10.1088/0034-4885/73/4/046901}{{\em Rept. Prog.
  Phys.} {\bfseries 73} (2010) 046901},
\href{http://arxiv.org/abs/0911.5004}{{\ttfamily arXiv:0911.5004 [gr-qc]}}.

\bibitem{Chakraborty:2019vki}
S.~Chakraborty, D.~Kothawala, and A.~Pesci, ``{Raychaudhuri equation with zero
  point length},''
\href{http://arxiv.org/abs/1904.09053}{{\ttfamily arXiv:1904.09053 [gr-qc]}}.

\bibitem{tHooft:1996rdg}
G.~'t~Hooft, ``{The Scattering matrix approach for the quantum black hole: An
  Overview},'' \href{http://dx.doi.org/10.1142/S0217751X96002145}{{\em Int. J.
  Mod. Phys.} {\bfseries A11} (1996) 4623--4688},
\href{http://arxiv.org/abs/gr-qc/9607022}{{\ttfamily arXiv:gr-qc/9607022
  [gr-qc]}}.

\bibitem{Sen:1995in}
A.~Sen, ``{Extremal black holes and elementary string states},''
  \href{http://dx.doi.org/10.1142/S0217732395002234}{{\em Mod. Phys. Lett.}
  {\bfseries A10} (1995) 2081--2094},
\href{http://arxiv.org/abs/hep-th/9504147}{{\ttfamily arXiv:hep-th/9504147
  [hep-th]}}.

\bibitem{Strominger:1996sh}
A.~Strominger and C.~Vafa, ``{Microscopic origin of the Bekenstein-Hawking
  entropy},'' \href{http://dx.doi.org/10.1016/0370-2693(96)00345-0}{{\em Phys.
  Lett.} {\bfseries B379} (1996) 99--104},
\href{http://arxiv.org/abs/hep-th/9601029}{{\ttfamily arXiv:hep-th/9601029
  [hep-th]}}.

\bibitem{Rovelli:1996dv}
C.~Rovelli, ``{Black hole entropy from loop quantum gravity},''
  \href{http://dx.doi.org/10.1103/PhysRevLett.77.3288}{{\em Phys. Rev. Lett.}
  {\bfseries 77} (1996) 3288--3291},
\href{http://arxiv.org/abs/gr-qc/9603063}{{\ttfamily arXiv:gr-qc/9603063
  [gr-qc]}}.

\bibitem{Ashtekar:1997yu}
A.~Ashtekar, J.~Baez, A.~Corichi, and K.~Krasnov, ``{Quantum geometry and black
  hole entropy},'' \href{http://dx.doi.org/10.1103/PhysRevLett.80.904}{{\em
  Phys. Rev. Lett.} {\bfseries 80} (1998) 904--907},
\href{http://arxiv.org/abs/gr-qc/9710007}{{\ttfamily arXiv:gr-qc/9710007
  [gr-qc]}}.

\bibitem{Balachandran:2013cq}
A.~P. Balachandran, T.~R. Govindarajan, A.~R. de~Queiroz, and A.~F. Reyes-Lega,
  ``{Algebraic Approach to Entanglement and Entropy},''
  \href{http://dx.doi.org/10.1103/PhysRevA.88.022301}{{\em Phys. Rev.}
  {\bfseries A88} no.~2, (2013) 022301},
\href{http://arxiv.org/abs/1301.1300}{{\ttfamily arXiv:1301.1300 [math-ph]}}.

\bibitem{Solodukhin:2011gn}
S.~N. Solodukhin, ``{Entanglement entropy of black holes},''
  \href{http://dx.doi.org/10.12942/lrr-2011-8}{{\em Living Rev. Rel.}
  {\bfseries 14} (2011) 8},
\href{http://arxiv.org/abs/1104.3712}{{\ttfamily arXiv:1104.3712 [hep-th]}}.

\bibitem{Solodukhin:2008dh}
S.~N. Solodukhin, ``{Entanglement entropy, conformal invariance and extrinsic
  geometry},'' \href{http://dx.doi.org/10.1016/j.physletb.2008.05.071}{{\em
  Phys. Lett.} {\bfseries B665} (2008) 305--309},
\href{http://arxiv.org/abs/0802.3117}{{\ttfamily arXiv:0802.3117 [hep-th]}}.

\bibitem{Govindarajan:2011hs}
T.~R. Govindarajan and R.~Tibrewala, ``{Novel black hole bound states and
  entropy},'' \href{http://dx.doi.org/10.1103/PhysRevD.83.124045}{{\em Phys.
  Rev.} {\bfseries D83} (2011) 124045},
\href{http://arxiv.org/abs/1102.4919}{{\ttfamily arXiv:1102.4919 [gr-qc]}}.

\bibitem{Govindarajan:2015hfa}
T.~R. Govindarajan and R.~Tibrewala, ``{Fermionic edge states and new
  physics},'' \href{http://dx.doi.org/10.1103/PhysRevD.92.045040}{{\em Phys.
  Rev.} {\bfseries D92} no.~4, (2015) 045040},
\href{http://arxiv.org/abs/1506.05243}{{\ttfamily arXiv:1506.05243 [hep-th]}}.

\bibitem{Govindarajan:2013xka}
T.~R. Govindarajan and V.~P. Nair, ``{Quantum field theories with boundaries
  and novel instabilities},''
  \href{http://dx.doi.org/10.1103/PhysRevD.89.025020}{{\em Phys. Rev.}
  {\bfseries D89} no.~2, (2014) 025020},
\href{http://arxiv.org/abs/1310.2034}{{\ttfamily arXiv:1310.2034 [hep-th]}}.

\bibitem{Deser:1997ri}
S.~Deser and O.~Levin, ``{Accelerated detectors and temperature in (anti)-de
  Sitter spaces},'' \href{http://dx.doi.org/10.1088/0264-9381/14/9/003}{{\em
  Class. Quant. Grav.} {\bfseries 14} (1997) L163--L168},
\href{http://arxiv.org/abs/gr-qc/9706018}{{\ttfamily arXiv:gr-qc/9706018
  [gr-qc]}}.

\bibitem{Jacobson:1997ux}
T.~Jacobson, ``{Comment on `Accelerated detectors and temperature in anti-de
  Sitter spaces'},'' \href{http://dx.doi.org/10.1088/0264-9381/15/1/020}{{\em
  Class. Quant. Grav.} {\bfseries 15} (1998) 251--253},
\href{http://arxiv.org/abs/gr-qc/9709048}{{\ttfamily arXiv:gr-qc/9709048
  [gr-qc]}}.

\bibitem{Deser:1998xb}
S.~Deser and O.~Levin, ``{Mapping Hawking into Unruh thermal properties},''
  \href{http://dx.doi.org/10.1103/PhysRevD.59.064004}{{\em Phys. Rev.}
  {\bfseries D59} (1999) 064004},
\href{http://arxiv.org/abs/hep-th/9809159}{{\ttfamily arXiv:hep-th/9809159
  [hep-th]}}.

\bibitem{Padmanabhan:2002ha}
T.~Padmanabhan, ``{Thermodynamics and / of horizons: A Comparison of
  Schwarzschild, Rindler and de Sitter space-times},''
  \href{http://dx.doi.org/10.1142/S021773230200751X}{{\em Mod. Phys. Lett.}
  {\bfseries A17} (2002) 923--942},
\href{http://arxiv.org/abs/gr-qc/0202078}{{\ttfamily arXiv:gr-qc/0202078
  [gr-qc]}}.

\bibitem{Banerjee:2010ma}
R.~Banerjee and B.~R. Majhi, ``{A New Global Embedding Approach to Study
  Hawking and Unruh Effects},''
  \href{http://dx.doi.org/10.1016/j.physletb.2010.05.001}{{\em Phys. Lett.}
  {\bfseries B690} (2010) 83--86},
\href{http://arxiv.org/abs/1002.0985}{{\ttfamily arXiv:1002.0985 [gr-qc]}}.

\bibitem{Santos:2004ws}
N.~L. Santos, O.~J.~C. Dias, and J.~P.~S. Lemos, ``{Global embedding of
  D-dimensional black holes with a cosmological constant in Minkowskian
  spacetimes: Matching between Hawking temperature and Unruh temperature},''
  \href{http://dx.doi.org/10.1103/PhysRevD.70.124033}{{\em Phys. Rev.}
  {\bfseries D70} (2004) 124033},
\href{http://arxiv.org/abs/hep-th/0412076}{{\ttfamily arXiv:hep-th/0412076
  [hep-th]}}.

\bibitem{Kim:2000ct}
Y.-W. Kim, Y.-J. Park, and K.-S. Soh, ``{Reissner-Nordstrom AdS black hole in
  the GEMS approach},''
  \href{http://dx.doi.org/10.1103/PhysRevD.62.104020}{{\em Phys. Rev.}
  {\bfseries D62} (2000) 104020},
\href{http://arxiv.org/abs/gr-qc/0001045}{{\ttfamily arXiv:gr-qc/0001045
  [gr-qc]}}.

\bibitem{Chen:2004qw}
H.-Z. Chen, Y.~Tian, Y.-H. Gao, and X.-C. Song, ``{The GEMS approach to
  stationary motions in the spherically symmetric spacetimes},''
  \href{http://dx.doi.org/10.1088/1126-6708/2004/10/011}{{\em JHEP} {\bfseries
  10} (2004) 011},
\href{http://arxiv.org/abs/gr-qc/0409107}{{\ttfamily arXiv:gr-qc/0409107
  [gr-qc]}}.

\bibitem{Banados:1992wn}
M.~Banados, C.~Teitelboim, and J.~Zanelli, ``{The Black hole in
  three-dimensional space-time},''
  \href{http://dx.doi.org/10.1103/PhysRevLett.69.1849}{{\em Phys. Rev. Lett.}
  {\bfseries 69} (1992) 1849--1851},
\href{http://arxiv.org/abs/hep-th/9204099}{{\ttfamily arXiv:hep-th/9204099
  [hep-th]}}.

\bibitem{Gannouji:2013eka}
R.~Gannouji and N.~Dadhich, ``{Stability and existence analysis of static black
  holes in pure Lovelock theories},''
  \href{http://dx.doi.org/10.1088/0264-9381/31/16/165016}{{\em Class. Quant.
  Grav.} {\bfseries 31} (2014) 165016},
\href{http://arxiv.org/abs/1311.4543}{{\ttfamily arXiv:1311.4543 [gr-qc]}}.

\bibitem{Dadhich:2012ma}
N.~Dadhich, J.~M. Pons, and K.~Prabhu, ``{On the static Lovelock black
  holes},'' \href{http://dx.doi.org/10.1007/s10714-013-1514-0}{{\em Gen. Rel.
  Grav.} {\bfseries 45} (2013) 1131--1144},
\href{http://arxiv.org/abs/1201.4994}{{\ttfamily arXiv:1201.4994 [gr-qc]}}.

\bibitem{gravitation}
T.Padmanabhan, {\em {Gravitation: Foundations and Frontiers}}.
\newblock Cambridge University Press, Cambridge, UK, 2010.

\bibitem{Wald:1999vt}
R.~M. Wald, ``{The thermodynamics of black holes},'' {\em Living Rev.Rel.}
  {\bfseries 4} (2001) 6,
\href{http://arxiv.org/abs/gr-qc/9912119}{{\ttfamily arXiv:gr-qc/9912119
  [gr-qc]}}.

\bibitem{Padmanabhan:2019yyg}
T.~Padmanabhan, ``{Thermality of the Rindler horizon: A simple derivation from
  the structure of the inertial propagator},''
\href{http://arxiv.org/abs/1905.08263}{{\ttfamily arXiv:1905.08263 [gr-qc]}}.

\bibitem{Rajeev:2019bzv}
K.~Rajeev and T.~Padmanabhan, ``{Exploring the Rindler vacuum and the Euclidean
  Plane},''
\href{http://arxiv.org/abs/1906.09278}{{\ttfamily arXiv:1906.09278 [gr-qc]}}.

\bibitem{Arzano:2018oby}
M.~Arzano and J.~Kowalski-Glikman, ``{Horizon temperature on the real line},''
  \href{http://dx.doi.org/10.1016/j.physletb.2018.10.019}{{\em Phys. Lett. B}
  {\bfseries C788} (2019) 82--86},
\href{http://arxiv.org/abs/1804.10550}{{\ttfamily arXiv:1804.10550 [hep-th]}}.

\end{thebibliography}\endgroup

\bibliographystyle{./utphys1}
\end{document}